\documentstyle[sprocl]{article}
\def\Journal#1#2#3#4{{#1} {\bf #2}, #3 (#4)}

\def\NPB{{\em Nucl. Phys.} B}
\def\PLB{{\em Phys. Lett.}  B}

\def\IJMPA{{\em Int. J. Mod. Phys.}A}
\def\CMP{\em Comm. Math. Phys.}

\def\be{\begin{equation}}
\def\ee{\end{equation}}
\def\bea{\begin{eqnarray}}
\def\eea{\end{eqnarray}}

\newcommand{\ft}[2]{{\textstyle\frac{#1}{#2}}}

\def\bigone{{\hbox{1\kern -.23em{\rm l}}}}
\newsavebox{\uuunit}
\sbox{\uuunit}
    {\setlength{\unitlength}{0.825em}
     \begin{picture}(0.6,0.7)
        \thinlines
        \put(0,0){\line(1,0){0.5}}
        \put(0.15,0){\line(0,1){0.7}}
        \put(0.35,0){\line(0,1){0.8}}
       \multiput(0.3,0.8)(-0.04,-0.02){12}{\rule{0.5pt}{0.5pt}}
     \end {picture}}

\newcommand{\Ka}{K\"ahler}

\newcommand  {\Rbar} {{\mbox{\rm$\mbox{I}\!\mbox{R}$}}}

\newcommand {\Cbar}
    {\mathord{\setlength{\unitlength}{1em}
     \begin{picture}(0.6,0.7)(-0.1,0)
        \put(-0.1,0){\rm C}
        \thicklines
        \put(0.2,0.05){\line(0,1){0.55}}
     \end {picture}}}
\def\IZ{{\hbox{{\rm Z}\kern-.4em\hbox{\rm Z}}}}
\def\Im{{\rm Im ~}}
\def\Re{{\rm Re ~}}
\newcommand{\Sp}[1]{\mbox{$Sp\left( #1,\Rbar \right) $}}
\newcommand{\symp}[1]{\mbox{$Sp\left( #1,\Rbar \right) $}}
\newcommand{\sinprod}[2]{\mbox{$\langle #1 , #2 \rangle$}}
\newcommand{\eqn}[1]{Eq.~\ref{#1}}

\begin{document}
\begin{titlepage}
\begin{flushright} KUL-TF-96/11 \\ hep-th/9606073
\end{flushright}
\vfill
\begin{center}
{\LARGE\bf The Definitions of Special Geometry} $^1$   \\
\vskip 27.mm  \large
{\bf   Ben Craps, Frederik Roose, \\[2mm] Walter Troost $^2$
and Antoine Van Proeyen $^3$} \\
\vskip 1cm
{\em Instituut voor theoretische fysica}\\
{\em Universiteit Leuven, B-3001 Leuven, Belgium}
\end{center}
\vfill

\begin{center}
{\bf ABSTRACT}
\end{center}
\begin{quote}
  The scalars in vector multiplets of N=2 supersymmetric theories in
  4 dimensions exhibit `special  K\"ahler geometry', related
  to  duality symmetries, due to their coupling to the vectors.
  In the literature there is some confusion on the definition of special
  geometry, which we want to clear up, for rigid as well as for local N=2.
\vfill      \hrule width 5.cm
\vskip 2.mm
{\small
\noindent $^1$ To be published in the proceedings of the second
international Sakharov conference on physics, Moscow, may 1996.
\\
\noindent $^2$ Onderzoeksleider NFWO, Belgium; E-mail:
Walter.Troost@fys.kuleuven.ac.be   \\
\noindent $^3$ Onderzoeksleider NFWO, Belgium; E-mail:
Antoine.VanProeyen@fys.kuleuven.ac.be}
\end{quote}
\begin{flushleft}
June 1996
\end{flushleft}
\end{titlepage}

\title{THE DEFINITIONS OF SPECIAL GEOMETRY }

\author{ B. CRAPS, F. ROOSE, W. TROOST AND A. VAN PROEYEN }

\address{Instituut voor theoretische fysica, Universiteit Leuven,\\
B-3001 Leuven, Belgium }

\maketitle\abstracts{
  The scalars in vector multiplets of N=2 supersymmetric theories in
  4 dimensions exhibit `special  K\"ahler geometry', related
  to  duality symmetries, due to their coupling to the vectors.
  In the literature there is some confusion on the definition of special
  geometry, which we want to clear up, for rigid as well as for local N=2.
  }

\section{Introduction}
In four dimensions, the duality transformations are transformations
between the field strengths of spin-1 fields. The kinetic terms of
the vectors (we only treat abelian vectors) are generically
\begin{equation}
{\cal L}_1=
\ft14 (\Im {\cal N}_{\Lambda\Sigma}){\cal F}_{\mu\nu}^\Lambda
{\cal F}^{\mu\nu\Sigma}
-\ft i8 (\Re {\cal N}_{\Lambda\Sigma})
\epsilon^{\mu\nu\rho\sigma}{\cal F}_{\mu\nu}^\Lambda
{\cal F}_{\rho\sigma}^\Sigma\ ,   \label{L1}
\end{equation}
where $\Lambda,\Sigma=1, ..., m$, and
${\cal N}_{\Lambda\Sigma}$  depend on the scalars. The
transformations of interest preserve the set of field equations and Bianchi
identities and form a group 
\symp{2m}, defined by matrices
\begin{equation}
{\cal S}=\pmatrix{A&B\cr C&D\cr} \qquad\mbox{where} \qquad
 {\cal S}^T  \Omega   {\cal S}   =  \Omega  \qquad\mbox{and}\qquad
\Omega=\pmatrix{0&\bigone \cr -\bigone &0\cr}\ .
\end{equation}
Under the action of this group the coupling constant matrix ${\cal
N}$ should also transform:
\begin{equation}
\tilde{\cal N} = (C + D{\cal N})(A+B{\cal N})^{-1}\ , \label{cNsympl}
\end{equation}
which demands a relation between the dualities and the scalar sector.

A relation between the scalars and the vectors arises naturally in $N=2$
supersymmetry vector multiplets (or larger $N$ extensions).
The geometry defined by the couplings of the complex scalars in the
supergravity version~\cite{DWVP}
has been given the name \cite{special} `special \Ka\ geometry'. The
similar geometry in rigid supersymmetry \cite{PKTN2} has been called
`rigid special \Ka'.

The above paragraph determines the concept of a special \Ka\ manifold.
Various authors
\cite{special,CdAF,CDFLL,f0art,prtrquat,fresoriabook,ItalianN2}
have proposed a definition which does not refer explicitly to
supersymmetry. However, these definitions are not completely
equivalent, as we will explain. We will give new definitions,
and have proven their equivalence.
Proofs and examples are worked out in detail in \cite{defspg}.
%
%

\section{Rigid special geometry}\label{ss:rigid}
The vector multiplets of $N=2$ supersymmetry
are constrained chiral superfields. The
construction of an action starts by introducing a holomorphic
function $F(X)$ of the lowest components of the multiplets, $X^A$,
where $A=1, ..., n$ labels the different multiplets. The kinetic
terms of the scalars then define a \Ka\ manifold with \Ka\ potential
\begin{equation}
K(z, \bar z)=i \left( X^A\frac{\partial}{\partial\bar X^A}\bar
F(\bar X)- \bar X^A \frac{\partial}{\partial X^A} F(X)\right)\ .
\label{Karigid}\end{equation}
The $X^A$ are functions of
$z^\alpha$ for $\alpha=1, ..., n$, which are arbitrary coordinates of the
scalar manifolds. The straightforward choice $z^\alpha = X^A$ is
called `special coordinates'.
The coupling of the $n$ vectors with the scalars is described as in
\eqn{L1} by the tensor
\begin{equation}
{\cal N}_{AB}= \frac{\partial}{\partial\bar X^A}
\frac{\partial}{\partial\bar X^B}\bar F\ . \label{defcNN2}
\end{equation}
Note that the positivity condition for the \Ka\ metric is the same as
the requirement of negative definiteness of $\Im{\cal N}_{AB}$, which
guarantees the correct sign for the kinetic energies of the vectors
(this condition is preserved by symplectic transformations
\eqn{cNsympl}).

One can check that \eqn{defcNN2} leads to the transformation \eqn{cNsympl}
for ${\cal N}$ if we combine $X^A$ and $F_A\equiv \partial F/\partial
X^A$ in a symplectic vector. Remark that then also the \Ka\ potential
\eqn{Karigid} gets the symplectic form
\begin{equation}
K= i \langle V,\bar V\rangle \qquad\mbox{where}\qquad
V=\pmatrix {X^A\cr F_A} \quad \mbox{and}\quad
\langle V , W \rangle \equiv V^T \Omega W \ . \label{Karigidsympl}
\end{equation}
The symplectic transformations do not always lead to the same action.
However, the scalar manifold remains always the same (the \Ka\
potential is invariant).

This leads to our {\bf first definition of a rigid special \Ka\
manifold.} It is a \Ka\ manifold\footnote{We always suppose here and
below that the metric is positive definite.} with on every chart
holomorphic functions $X^A(z)$ and a holomorphic function $F(X)$
such that the \Ka\ potential can be written as in \eqn{Karigid}.
Where charts $i$ and $j$ overlap, there should be transition
functions:
\begin{equation}
\left( \begin{array}{c}
 X \\ \partial F \end{array}\right)_{(i)} = e^{ic_{ij}} M_{ij}
\left( \begin{array}{c}
 X \\ \partial F\end{array}\right)_{(j)}+ b_{ij}\ ,  \label{ISpn}
\end{equation}
with $c_{ij}\in \Rbar$; $M_{ij} \in \symp{2n}$ and $b_{ij} \in
\Cbar^{2n}$. These overlap functions should satisfy the cocycle
condition on overlaps of 3 charts.

The previous definition makes use of the prepotential function
$F(X)$, which is not a symplectic invariant object. In the  second
definition, inspired by Strominger's definition \cite{special} for
local special geometry, we only use objects which have a meaning in
the symplectic bundle.

{\bf Definition 2:} A \Ka\ manifold ${\cal M}$ such that there
exists a complex U(1)-line bundle ${\cal L}$ with constant
transition functions, and an $I\Sp{2n}$-vector bundle (in the sense
of \eqn{ISpn}, thus with complex inhomogeneous part) ${\cal H}$ over
${\cal M}$. It should have a holomorphic section $V$ of ${\cal
L}\otimes{\cal H}$, such that the \Ka\ form is given by
\eqn{Karigidsympl}, and
\begin{equation}
\sinprod{\partial_\alpha V}{\partial_{\beta} V} = 0\ . \label{cond3def2}
\end{equation}

The last equation has no equivalent in \cite{special}. To argue
that it is appropriate, define
\begin{equation}
U_\alpha=\partial_\alpha V=\partial_\alpha\pmatrix{X^A\cr F_A}=
\pmatrix{e_\alpha^A\cr h_{\alpha A}}\ .
\end{equation}
Then $g_{\alpha\bar \beta}=i \sinprod {U_\alpha}{\bar U_{\bar
\beta}}$. The positivity requirement for this metric implies
invertibility of $e_\alpha^A$. Then the coupling constant matrix
${\cal N}$, \eqn{defcNN2}, can be written as $
\bar {\cal N}_{AB}\equiv \left( e_\alpha^A\right) ^{-1} h_{\alpha
B}$. The condition \eqn{cond3def2} is the requirement for the matrix
${\cal N}$ to be symmetric, and is therefore necessary.
In turn it implies the integrability
$F_A=\partial_A F(X)$, and thus the existence of the function $F$,
which is essential in the first definition.

The {\bf third definition} is convenient for defining special geometry in
the moduli space of Riemann surfaces. It starts from a manifold
with in each chart $2n$ closed 1-forms $U_\alpha$:
\begin{equation}
\partial_{\bar \alpha}U_{\beta}=0\ ;\qquad
\partial_{\left[\alpha\right.}U_{\left. \beta\right]}= 0
\label{Uclosed}\ ,
\end{equation}
such that
\begin{equation}
\sinprod{U_\alpha}{U_{\beta}}= 0 \ ;\qquad
\sinprod{U_\alpha}{\bar U_{\bar \beta}}=-ig_{\alpha\bar \beta}\ .
\label{Umetric}
\end{equation}
The first of equations \ref{Umetric} replaces \eqn{cond3def2}, while
the second one
defines the metric in this approach. In the transition
functions, the inhomogeneous term
(see \eqn{ISpn}) is no longer present. It reappears
in the integration from $U_\alpha$ to $V$ by \eqn{Uclosed}.

Rigid special geometries can be constructed by considering
Riemann surfaces parametrised by $n$ complex moduli. The $U_\alpha$
are then identified with integrals of $n$ holomorphic 1-forms
over a $2n$-dimensional canonical homology basis of 1-cycles. Most
of the equations of the third definition are then automatically
satisfied. The non-trivial one is the second one of \eqn{Uclosed}.

\section{(Local) Special geometry}\label{ss:local}
The name special geometry has been reserved \cite{special} for the
manifold determined by the scalars of vector multiplets in $N=2$
supergravity. The first construction~\cite{DWVP} of these actions
made use of superconformal tensor calculus. This setup, which
involves a local dilatation and an extra $U(1)$ gauge symmetry, gave
insight in the structure of these manifolds. E.g. the (auxiliary)
gauge field of the $U(1)$ turns into the connection of the \Ka\
transformations $K(z, \bar z) \rightarrow K(z, \bar z) + f(z) + \bar
f(\bar z)$. Also, this construction
involves an extra vectormultiplet (labelled with $0$), whose scalar
fields $X^0$ can be gauge fixed by this dilatation and extra $U(1)$,
while its vector becomes the physical graviphoton. This structure
with $(n+1)$ vectormultiplets with scaling invariance is at the basis
of the formulations below.

The first construction started (as in rigid susy) from chiral multiplets,
and thus leads in the same way to a holomorphic prepotential $F(X^I)$,
where $(I=0,1,...,n)$. But because of the scaling symmetry, there is
now an extra requirement: $F(X)$ should be homogeneous of weight 2,
where $X$ has weight~1. The $X^I$ thus span an $n+1$ dimensional
complex projective space. There are only $n$ complex physical
scalars, an overall factor is not relevant. One can choose
coordinates $z^\alpha $, with $\alpha=1,...,n$, such that the $X^I$
are proportional to $n+1$ holomorphic functions of these:
$X^I\propto  Z^I(z^\alpha)$.
One can then choose a gauge condition for the dilatations. To
decouple the kinetic terms of the scalars and the graviton one
chooses:
\begin{equation}
i(\bar X^I F_I -\bar F_I X^I)=i\langle \bar V , V\rangle =1
\qquad\mbox{with}\qquad
 V\equiv\pmatrix{X^I\cr F_J} \ ,
\end{equation}
and $F_I\equiv \partial F(X)/\partial X^I$. One obtains an
action where the scalars describe a \Ka\ manifold and
$V=e^{K/2}\, v(z)$, where  $v$
is holomorphic (its upper components are $Z^I(z)$, see above, and
the lower are $F_I(Z(z))$)
and $K$ is the \Ka\ potential
\begin{equation}
K(z,\bar z)= -\log\left[i
\bar Z^I \frac{\partial}{\partial Z^I} F(Z) -i
Z^I\frac{\partial}{\partial\bar Z^I}\bar F(\bar Z)
 \right]=
-\log\left[i \langle \bar v(\bar z) , v(z)\rangle \right]\ . \label{Kalocal}
\end{equation}

The {\bf first definition} of a special manifold then starts from
an $n$-dimensional Hodge-\Ka\ manifold with on any chart  complex
projective coordinates $Z^I(z)$, and a holomorphic function $F(Z^I)$
homogeneous of second degree, such that the \Ka\ potential is given
by \eqn{Kalocal}. On overlaps of charts $i$ and $j$, one should have
\begin{equation}
\left( \begin{array}{c}
 Z \\ \partial F \end{array}\right)_{(i)} = e^{f_{ij}(z)} M_{ij}
\left( \begin{array}{c}
 Z \\ \partial F\end{array}\right)_{(j)}\ ,
\end{equation}
with  $ f_{ij}$ holomorphic and  $M_{ij} \in \symp{2n+2}$.
 The overlap functions satisfy the cocycle condition on overlaps
of 3 charts.

Note that local special geometry involves local holomorphic
transition functions. This is related to the presence of the gauge
field of the local $U(1)$ in the superconformal approach.

The {\bf second definition} is, as in the rigid case, an intrinsic
symplectic formulation. It starts from an $n$-dimensional Hodge-\Ka\
manifold ${\cal M}$ with ${\cal L}$ a complex line bundle whose first
Chern class equals the \Ka\ form ${\cal K}$. There is a
$\symp{2(n+1)}$ vector bundle ${\cal H}$ over ${\cal M}$, and a
holomorphic section $v(z)$ of ${\cal L}\otimes{\cal H}$, such that the
\Ka\ potential is given by the last expression in \eqn{Kalocal}, and
such that
\begin{equation}
\sinprod{v}{\partial_{\alpha} v} = 0\ ;\qquad
\sinprod{\partial_\alpha v}{\partial_{\beta} v} = 0\ . \label{cond3local}
\end{equation}

This definition is a rewriting of Strominger's \cite{special}
definition, where, however, \eqn{cond3local} was missing. This
condition is necessary to have a symmetric matrix ${\cal N}_{IJ}$
(see \cite{f0art} for the definition of ${\cal N}$ in this context).
We have constructed counterexamples \cite{defspg} to show the
necessity of both conditions in general.
Actually, for $n>1$ the first one is a consequence of the
other constraints. For $n=1$ the second condition is empty.

In \cite{f0art} it was shown that
the prepotential $F(X)$ may not exist\footnote{ This case includes
some
physically important theories.}. Therefore one might wonder about the
equivalence of the two definitions. It has now been proven that in
all such cases there exists a symplectic transformation to a form
where $F$ exists~\cite{defspg}.
Although this does not invalidate the necessity of
the formulation without a prepotential for physical purposes (e.g.
to have actions invariant under larger parts of the isometry group
and gaugings thereof), it implies that to describe the scalar
manifolds, one can choose in each chart
such a symplectic basis. This is the essential step in proving the
equivalence of the two definitions.

There is a third definition (inspired by
\cite{CDFLL,f0art,prtrquat}),
which is analogous to the third definition in the rigid case. It
can be formulated using the $2(n+1)\times 2(n+1)$ matrix built from
the symplectic vectors $V,\,\bar V,\, U_\alpha,\, \bar U_{\bar \alpha}$
(the latter two now being defined by covariant derivatives on the first
two involving also the $U(1)$ connection). This formulation is then
appropriate for expressing the moduli space of Calabi-Yau surfaces as
a special manifold. This matrix is then identified with the
integrals of $(3,0), \, (2,1), \, (1,2)$ and $(0,3)$ forms over 3-cycles.
\section{Remark and conclusions}\label{ss:concl}
Special geometry is sometimes
defined by giving the curvature formulas.
Whereas these equations are always valid, we have not investigated
(and know of no proof elsewhere)
in how far they constitute a sufficient condition.

We have given several equivalent definitions of special geometry.
It is clear that the symplectic transformations, inherited from the
dualities on vectors, are crucial for the scalar manifolds.
We have seen that some equations are missing in nearly all earlier
proposals.
The missing equations are mostly related to the requirement of
symmetry~\cite{prtrquat} of the matrix ${\cal N}_{IJ}$,
which is a necessary
requirement for the discussion of symplectic transformations.

\noindent {\bf Acknowledgments}

\noindent
W.T. and A.V.P. are Senior research associate of the NFWO, Belgium.
This work was carried out in the framework of the
project "Gauge theories, applied supersymmetry and quantum
gravity", contract SC1-CT92-0789 of the European Economic
Community.

\end{document}